\documentclass[sigplan,nonacm]{acmart}
\usepackage{mathpartir}
\usepackage{epigraph}
\usepackage{bbding} 
\usepackage{xspace}
\usepackage{listings}
\usepackage{xcolor}
\usepackage{expkv-cs}
\usepackage{caption}

\newcommand\rtri{\triangleright}
\newcommand\pwd[2]{\mathrm{while}\ #1\ \mathrm{do}\ #2}

\newcommand\ite[3]{\mathrm{if}\ #1\ \mathrm{then}\ #2\ \mathrm{else}\ #3}

\newcommand\tc{\vdash}
\newcommand\CT{\textit{CT}}
\newcommand\CL{\textit{CL}}
\newcommand\Bool{\textbf{Bool}}
\newcommand\Int{\textbf{Int}}
\newcommand\true{\textbf{true}}
\newcommand\false{\textbf{false}}
\newcommand\classdecl[2]{\text{class}\ #1\ \text{extends}\ #2}
\newcommand\Obj{\textbf{Obj}}
\newcommand\GroundType{\mathcal{G}}
\newcommand\Type{\mathcal{T}}
\newcommand\MethodType{\mathcal{M}}

\newcommand{\quoting}[1]{``#1''}

\definecolor{lightgray}{gray}{0.9}

\definecolor{ShadowColor}{RGB}{30,150,190}

\makeatletter
\newcommand\Cshadowbox{\VerbBox\@Cshadowbox}
\def\@Cshadowbox#1{%
  \setbox\@fancybox\hbox{\fbox{#1}}%
  \leavevmode\vbox{%
    \offinterlineskip
    \dimen@=\shadowsize
    \advance\dimen@ .5\fboxrule
    \hbox{\copy\@fancybox\kern.5\fboxrule\lower\shadowsize\hbox{%
      \color{ShadowColor}\vrule \@height\ht\@fancybox \@depth\dp\@fancybox \@width\dimen@}}%
    \vskip\dimexpr-\dimen@+0.5\fboxrule\relax
    \moveright\shadowsize\vbox{%
      \color{ShadowColor}\hrule \@width\wd\@fancybox \@height\dimen@}}}
\makeatother

\newcommand{\ba}{\begin{array}}
\newcommand{\ea}{\end{array}}

\advance \topmargin by -\headheight
\advance \topmargin by -\headsep
\evensidemargin \oddsidemargin
\usepackage[T1]{fontenc}
\usepackage{graphicx}
\usepackage{times}

\usepackage{url}

\usepackage{listings}

\newcommand{\topic}[1]{}

\newcommand{\Cpp}{C\kern-0.05em\texttt{+\kern-0.03em+}}

\definecolor{lightgray}{gray}{0.9}

\newcommand{\key}[1]{\ensuremath{\mathtt{#1}}}

\lstdefinestyle{basic}{showstringspaces=false,columns=fullflexible,language=C++,escapechar=@,xleftmargin=1pc,%
basicstyle=\scriptsize\ttfamily,
commentstyle=\mdseries,
morekeywords={concept,model,require,where},
}

\makeatletter

\newcount\@tempcntc
\def\@citex[#1]#2{\if@filesw\immediate\write\@auxout{\string\citation{#2}}\fi
  \@tempcnta\z@\@tempcntb\m@ne\def\@citea{}\@cite{\@for\@citeb:=#2\do
    {\@ifundefined
       {b@\@citeb}{\@citeo\@tempcntb\m@ne\@citea\def\@citea{,}{\bf ?}\@warning
       {Citation `\@citeb' on page \thepage \space undefined}}%
    {\setbox\z@\hbox{\global\@tempcntc0\csname b@\@citeb\endcsname\relax}%
     \ifnum\@tempcntc=\z@ \@citeo\@tempcntb\m@ne
       \@citea\def\@citea{,}\hbox{\csname b@\@citeb\endcsname}%
     \else
      \advance\@tempcntb\@ne
      \ifnum\@tempcntb=\@tempcntc
      \else\advance\@tempcntb\m@ne\@citeo
      \@tempcnta\@tempcntc\@tempcntb\@tempcntc\fi\fi}}\@citeo}{#1}}
\def\@citeo{\ifnum\@tempcnta>\@tempcntb\else\@citea\def\@citea{,}%
  \ifnum\@tempcnta=\@tempcntb\the\@tempcnta\else
   {\advance\@tempcnta\@ne\ifnum\@tempcnta=\@tempcntb \else \def\@citea{--}\fi
    \advance\@tempcnta\m@ne\the\@tempcnta\@citea\the\@tempcntb}\fi\fi}

\definecolor{dkgreen}{rgb}{0,0.6,0}
\definecolor{gray}{rgb}{0.5,0.5,0.5}
\definecolor{mauve}{rgb}{0.58,0,0.82}
\definecolor{lightgray}{gray}{0.9}
\definecolor{darkergrey}{rgb}{0.75, 0.75, 0.75}

\newcommand{\leaveout}[1]{}

\newcommand{\purplSyntax}[1]{{\fontfamily{ugq}\selectfont{\textcolor{violet}{#1}}}}
\newcommand{\PurPL}{{\purplSyntax{PurPL}}\xspace}

\newcommand*{\ddefeq}{\mathrel{\vcenter{\baselineskip0.5ex \lineskiplimit0pt
                     \hbox{\scriptsize.}\hbox{\scriptsize.}}}%
                      \mathrel{\vcenter{\baselineskip0.5ex \lineskiplimit0pt
                     \hbox{\scriptsize.}\hbox{\scriptsize.}}}
                     =\,}
\mathchardef\mhyphen="2D
\ekvcSplit\stctx{ct={\CT}, c={C}, g={\Gamma}, d={\Delta}}{#1; #2; #3; #4}
\newcommand*{\defeq}{\mathrel{\vcenter{\baselineskip0.5ex \lineskiplimit0pt
                     \hbox{\scriptsize.}\hbox{\scriptsize.}}}%
                     =}

\usepackage{etoolbox}
\AtBeginEnvironment{itemize}{%
  \setlength{\itemsep}{1.8pt}%
  \setlength{\parsep}{1.8pt}%
  \setlength{\topsep}{1.8pt}%
}
\AtBeginEnvironment{enumerate}{%
  \setlength{\itemsep}{1.8pt}%
  \setlength{\parsep}{1.8pt}%
  \setlength{\topsep}{1.8pt}%
}

\setcopyright{none}
\acmISBN{}
\acmDOI{}

\title{A Typestate Approach to Purpose-aware Programming}

\author{Joan Montas}
\affiliation{
  \institution{University of Massachusetts Lowell}           
  \country{USA}                    
}

\author{Samuel Dodson}
\affiliation{
  \institution{University of Massachusetts Lowell}            
  \country{USA}                    
}

\author{Anitha Gollamudi}
\affiliation{
  \institution{University of Massachusetts Lowell}            
  \country{USA}                    
}

\author{Matteo Cimini}
\affiliation{
  \institution{University of Massachusetts Lowell}            
  \country{USA}                    
}

\begin{document}
\begin{abstract}

Real-world applications often require verification that sensitive data is being used for their intended purpose. However, existing literature offers limited results regarding compile-time guarantees in this domain. In this paper, we explore the use of typestate to reason about the purpose of data. 
In typestate, types have a state, which can transition to other states in the style of automata. 
In our approach, the state of the type of sensitive data is defined as the set of purposes for which the data can be used. This set can grow or shrink at runtime as purposes can be added or removed during execution.

In this paper, we have developed \PurPL, an object-oriented programming language that features a typestate system that is capable of reasoning about purposes and data usage compliance according to them. 

We give an overview of \PurPL~ through examples and 
present a formal type system. 
We have also implemented \PurPL's type checker, and we report on our experiments with type checking various programming scenarios that handle sensitive data. 

\end{abstract}
\maketitle

\section{Introduction}

Real-world applications often need to analyze whether data is being used for the purpose they have been intended. This is regulated by law in some cases, as in the examples of privacy laws such as GDPR~\cite{GDPR}, CCPA~\cite{ccpa}, HIPAA~\cite{hipaa}, the Privacy Act in Australia~\cite{priv_aus}, and the Protection of Personal Information in Japan, to name a few. 
In this context, the concepts of \emph{purpose} and \emph{consent} emerge as fundamental elements that software systems need to handle.  
To make an example, GDPR contains the following article. 

\epigraph{
\emph{[sic]} Personal data shall be collected for specified, explicit and legitimate \textbf{purposes} and not further processed in a manner that is incompatible with those purposes.
}{Article (5) of GDPR}

Consent must be explicitly granted for some data to be used for some purpose. For example, GDPR contains the following article. 

\epigraph{
\emph{[sic]} \textbf{Consent} should cover all processing activities carried out for the same purpose or purposes. When the processing has multiple purposes, consent should be given for all of them.
}{Article (32) of GDPR}



Compliance with these laws is a paramount concern for software companies. 
The Cambridge Analytica scandal~\cite{analytica} involving Facebook provided a stark illustration of the complexities associated with GDPR non-compliance. 
Non-compliance with privacy laws led to steep fines in notorious cases against Google \cite{googleCase} and Whatsapp \cite{whatsappCase}, among others. However, a major problem with privacy compliance is that software developers are left with little to no automated tools to reason about the data involved in their operations.


In order to enforce the aforementioned privacy regulations, software companies are required to perform analyses that are not typical in software development: 
\emph{For what purpose does the function \texttt{f} use a specific argument? 
Has the data that function \texttt{f} handle been declared for such a purpose? 
}
To exacerbate the situation further, the purpose of data can change over time during the execution of programs, as users have the right to grant or withdraw consent virtually at any time. 


Literature offers various solutions for checking consent violations at runtime but offers limited results insofar compile-time guarantees is concerned. 
(We will review related work in Section \ref{sec:related}.)


In this paper, we argue that \emph{typestate} \cite{Strom1983,Garcia2014}. is a natural approach for reasoning about the purpose of data at compile-time. To recall, a classic example of typestate is that of a type \key{File} that is equipped with a state that can be either \key{open} or \key{closed}. 
Intuitively, the state of the type follows a transition system. When a \key{read} operation is performed on a variable \key{f}, not only the type system checks that the type of \key{f} is \key{File}, but also that this type is in state \key{open}, and after the operation, the state of the type is still \key{open}. Similarly, when a \key{close} operation is performed on \key{f}, the type system requires that the type of \key{f} is in state \key{open}, and \key{closed} after the operation. 

The central idea of our solution is to use typestate to assign a set of purposes as state. Given some data, its type records the set of purposes for which the data can be used. 
This set can grow and shrink over time, as purposes can be granted and revoked at runtime. 
Our examples show that this treatment is naturally accompanied with \emph{row-polymorphism}. Intuitively, methods can request specific purposes and bind \quoting{all other purposes} in order to enforce that these latter are present in other types.


We observe that purposes themselves have a state. Some purpose may not be activated yet or may be suspended for a time. The ability to reason about this aspect provides great expressiveness: Methods declarations can specify that they can be invoked only so long as some purpose is in a certain state, and that the state of a purpose changes after the execution of the method.

In this paper, we present our ongoing work to develop \PurPL (as in \purplSyntax{Pur}pose-aware \purplSyntax{P}rogramming \purplSyntax{L}anguage), 
an object-oriented programming language with a typestate system that accommodates the features just discussed. \PurPL~ uses typestate nontrivially for 1) assigning purpose sets as states, and row-polymorphism with it, and 2) adding a \quoting{state to states}.


This paper makes the following contributions. 

\begin{itemize}
\item We provide an overview of \PurPL~ through a handful of examples. 
\item We present a formal type system that accommodates the features of \PurPL. 

\item We have implementated of our type system. 

\item We evaluate our type checker in the context of a handful of examples that test common dynamics with sensitive data and purpose. 

\item We delineate our ongoing work to make \PurPL~ a practical programming language. 

\end{itemize}
Ultimately, our type system focuses on enforcement of purpose compliance, providing a way to express various scenarios involving sensitive data and consent. It is important to note that the paper does not focus on Information Flow Control (IFC), which remains outside of the scope of this paper. However, we acknowledge that the addition of an IFC methodology is not negotiable to achieve defense-in-depth.

\paragraph{
Structure of the paper}
Section \ref{sec:overview} provides an overview of our system through examples. 
Sections \ref{sec:syntax} and \ref{sec:addingState} present a formalization of our typestate system. 
Section \ref{sec:syntax} addresses \PurPL without states for purposes (active, suspended, etcetera). Section \ref{sec:addingState} adds that feature to our formalism. 
Section \ref{sec:evaluation} describes our implementation and reports on our evaluation. 
Section \ref{sec:future} discusses the limitations of our system and future work. 
Section \ref{sec:related} discusses related work and  
Section \ref{sec:conclusion} concludes the paper. 

\section{Overview of Approach}\label{sec:overview}


Consider the scenario of a patient \texttt{P} that has recently received a treatment from a hospital. Upon this occurrence, the hospital stores various information about the patient \texttt{P}. 
These are administrative and demographic information that include \texttt{P.id} (a unique identifier of the patient) and patient's email address  \texttt{P.emailAddress}. 
The hospital also stores information about the patient's physical condition which includes their  obesity status \texttt{Phys.obesity}. 
Information about infectious diseases are also stored, among which there is \texttt{Inf.std} to records the status of the patient w.r.t. STDs. 
We assume that \texttt{Phys.obesity} and \texttt{Inf.std} are integers (an internal numeric code). 

In \PurPL, these variables have been declared to be used for some specific purpose. 

\paragraph{Purpose Compliance}
For our first example, let us suppose that the hospital wants to send a survey questionnaire to the email address of the patient \texttt{P}. 
Let us suppose that \texttt{P.email} has been granted for the purpose \texttt{Survey}, that is, the email address of the patient can be used by the hospital to send the survey. The following code models this scenario. 

{
\small
\begin{lstlisting}[numbers=left,frame=single, xleftmargin=15pt,  caption={\strut}, label={lst:one}]
Class Hospital {
void sendSurvey(email : string {| Survey |}) 
    {
    ... code ...
    }
}
H : Hospital := new Hospital(); 
P.email : string {| Survey |} := (*@\textit{email}@*);
H.sendSurvey(P.email); (*@ \textcolor{dkgreen}{\CheckmarkBold}} 
\end{lstlisting}
}

Line 8 declares the variable \texttt{P.emailAddress} for the purpose \texttt{Survey}. 
Lines 2-5 declare \texttt{sendSurvey} to have the argument \texttt{email}. Furthermore, this method requests that this argument \emph{must} be declared for \texttt{Survey}. At compile-time, our type system checks that this is the case, and therefore the function call at line 9 type checks successfully. 

\paragraph{Granting Purpose}

Suppose our patient wants to enroll in a research study that is related to obesity with a research institution \texttt{RI}. The patient expresses the desire to share its obesity status with \texttt{RI}. 
By mistake, however, \texttt{RI} requests \texttt{P}'s std status, which should not be sent to \texttt{RI}. 
\PurPL~ models this scenario as follows. \\

{\small
\begin{lstlisting}[numbers=left,frame=single, 
xleftmargin=16pt,
caption={\strut}, label={lst:two}]
Class ResearchInst {
void enroll(id : int {| RI_Trial |}, 
             obs_arg : int {| RI_Trial |})  {
    ... code ...
    }
}
... code, Phys.obesity (*@\textit{has purpose }@*)PhysExam
RI : ResearchInst = new ResearchInst();
P.id.grant(RI_Trial);
Phys.obesity.grant(RI_Trial);
RI.enroll(P.id, Inf.std); (*@ \textcolor{red}{\XSolidBrush}
\end{lstlisting}
}

(Notice that \texttt{Phys.obesity} and \texttt{Inf.std} have been previously defined, as explained above.)  
Lines 2-5 contain the declaration of the function \texttt{enroll}. This function requests the id of a patient and its obesity status and both must be declared for the purpose of this specific study, that is, purpose \texttt{RI\_Trial}. Lines 9 and 10 grant that information to be used for such a purpose thanks to \PurPL's primitive operation \texttt{grant}. 
Line 11 is type rejected due to the mistaken argument to \texttt{enroll}. In particular, \texttt{Inf.sdt} has not been declared for the purpose \texttt{RI\_Trial}. 
In this case, \PurPL's reasoning over purposes provides great protection on the patient's data. 

Once the error has been discovered, we can quickly fix the call at line 11 as follows.

{\small
\begin{lstlisting}[numbers=left,firstnumber=11, 
xleftmargin=16pt,
frame=single]
RI.enroll(P.id, Phys.obesity); (*@ \textcolor{red}{\XSolidBrush}
\end{lstlisting}
}

Except that the invocation is still type rejected. 
The reason for this is that line 2 and 3 require the two arguments to exactly have a singleton set of purposes with \texttt{RI\_Trial} only, whereas \texttt{Phys.obesity} also contains \texttt{PhysExam}.

\paragraph{Row Polymorphism over Purposes} 

Our type system should allow for \quoting{other purposes} to be present also. The main mechanism for doing this is subtyping. However, we show in later examples that it is convenient in our setting to \emph{bind} these \quoting{other purposes} and use them to specify sets of purposes in other places. We therefore employ \emph{row polymorphism} over sets of purposes.

To illustrate this feature, consider the following modification to the function declaration of \texttt{enroll}. 

{\small
\begin{lstlisting}[numbers=left,frame=single, 
xleftmargin=16pt,
caption={\strut}, label={lst:three}]
Class ResearchInst {
void enroll(
    id: int {| RI_Trial | rho1 |}, 
    obs_arg : int {| RI_Trial  | rho2 |})  {
    ... code ...
    }
}
... code, Phys.obesity (*@\textit{has purpose }@*)PhysExam
RI : ResearchInst = new ResearchInst();
P.id.grant(RI_Trial);
Phys.obesity.grant(RI_Trial);
RI.enroll(P.id, Phys.obesity); (*@ \textcolor{dkgreen}{\CheckmarkBold}@*)
\end{lstlisting}
}

\PurPL is able to express that the set of purposes of the argument \texttt{obs\_arg}, for example, must match \texttt{\{| RI\_Trial  | rho2 |\}}. 
The variable \texttt{P.obesity}, with purposes \texttt{\{| RI\_Trial, PhysExam |\}}, matches that request with \texttt{rho2} = \texttt{\{| PhysExam |\}}. Line 12 type checks successfully.

Notice that both arguments \texttt{id} and \texttt{obs\_arg} could use the same row variable, say \texttt{rho}, rather than  \texttt{rho1} and \texttt{rho2}. In that case, \PurPL~ checks that the rest of the purposes are also the same for the two. 
This is however not a sensible choice in the example above.

\paragraph{Reasoning with Row Variables} 

The presence of row variables such as \texttt{rho1} and \texttt{rho2} poses some challenges. 
For example, the code of \texttt{enroll} needs to handle a variable \texttt{obs\_arg} whose type now contains that variable, and \texttt{obs\_arg} may be passed to methods that themselves use row variables. Yet, our type system needs to determine whether it is safe to pass \texttt{obs\_arg}. 
To make an example, let us consider the following implementation of \texttt{enroll}. 

{\small
\begin{lstlisting}[numbers=left,frame=single, 
xleftmargin=16pt, caption={\strut}, label={lst:four}]
void press_release(
   id : int {| Press | rho |}) {
   ... code ...
    }

void enroll(
   id: int {| RI_Trial | rho1 |}, 
   obs_arg : int {| RI_Trial  | rho2 |})  {
      ... code ...
      press_release(id); (*@ \textcolor{red}{\XSolidBrush} @*)
    }
... code, (*@\textit{Lines 7-10 above}@*)
P.id.grant(Press);
RI.enroll(P.id, Phys.obesity); 
\end{lstlisting}
}

Here, \texttt{enroll} calls the method \texttt{press\_release} passing \texttt{id}. 
A part of the information that is publicly released is the unique identifiers of participants of the trial. 
The method \texttt{press\_release} requires the purpose \texttt{Press} for its argument. 
Unfortunately, the method called at line 10 cannot determine at compile-time whether the formal argument \texttt{id} has purpose \texttt{Press}. 
It may be one of the purposes in \texttt{rho1} but \PurPL cannot make the assumption that it is there. 
Therefore, line 10 is type rejected. 
\PurPL handles this with a pattern matching operation between two sets of purposes that can potentially both have row variables. 
In this case, \texttt{\{| Press | rho |\}} is not a pattern of \texttt{\{| RI\_Trial | rho1 |\}}. 

To solve this problem, \texttt{enroll} needs to know that the argument \texttt{id} is also for the \texttt{Press} purpose, as folllows. 

{\small
\begin{lstlisting}[numbers=left,frame=single,
xleftmargin=15pt,
caption={\strut}, label={lst:five}]
void enroll(
    id: int {| RI_Trial, Press | rho1 |}, 
    obs_arg : int {| RI_Trial  | rho2 |})  {
    ... code ...
    press_release(id); (*@ \textcolor{dkgreen}{\CheckmarkBold} @*)
    }
... code, (*@\textit{Lines 7-10 above}@*)
P.id.grant(Press);
RI.enroll(P.id, Phys.obesity); (*@ \textcolor{dkgreen}{\CheckmarkBold} @*)
\end{lstlisting}
}

In this case, \texttt{\{| Press | rho |\}} \emph{is} a pattern that accommodates \texttt{\{| RI\_Trial, Press | rho1 |\}} when \texttt{rho} = \texttt{\{| RI\_Trial | rho1 |\}}. 
Line 5 above type checks successfully. 

\paragraph{Purpose Change after Method Invocation} 

We show an example where method arguments update their purposes after the method has been called. Suppose that the method \texttt{end\_trial} in the class \texttt{ResearchInst} ends the trial for a specific patient. Next to the various closing operations, the method also removes the purposes that are related to the study from its arguments. 
The following code models this scenario. 

{\small
\begin{lstlisting}[numbers=left,frame=single, 
xleftmargin=16pt,
caption={\strut}, label={lst:six}]
void end_trial(
  id: int {| RI_Trial, Press | rho1 |} 
            => {| rho1 |}, 
  obs_arg : int {| RI_Trial  | rho2 |} 
            => {| rho2 |})  {
    ... code ...
    P.id.revoke(RI_Trial, Press);
    Phys.obesity.revoke(RI_Trial);
    }
    
... (*@\textit{previous code that enrolls the patient}@*)

RI.end_trial(P.id,Phys.obesity); 
RI.enroll(P.id, Phys.obesity); (*@ \textcolor{red}{\XSolidBrush} @*)
\end{lstlisting}
}

Notation such as \texttt{=> \{| rho1 |\}} of line 3 is borrowed from other typestate systems \cite{Garcia2014}. This notation says that the purposes bound to \texttt{rho1} are the set of purposes for the actual argument given for \texttt{id} after the method has been executed. Notably, these purposes do not contain \texttt{RI\_Trial}and \texttt{Press}. (Analogous treatment is specified for \texttt{obs\_arg}.)  

After line 13 is executed, the set of purposes of \texttt{P.id} and \texttt{Phys.obesity} do not have purposes related to the trial. 
Suppose that this trial allows a patient to conduct the study only once. When we mistakenly try to enroll our patient a second time at line 14, the call to \texttt{RI.enroll} is type rejected because its arguments do not have the required purposes. 
In this case, \PurPL protected the integrity of the trial (at compile-time) by ensuring that the patient's data could no longer be used for the trial's purposes. 

In order to type check successfully, the body of  the method \texttt{end\_trial} must support the state change declared in the signature at lines 3 and 5. This can be done using, for example, \PurPL's primitive operation \texttt{revoke}, which revokes a purpose from the set of purposes of data. 

\paragraph{Purposes Themselves Have a State} 

Some purposes are not valid at the beginning of a program because they become available only after some time. Similarly, some purposes may be suspended for a time. 
In short, purposes have a state. \PurPL~ provides the following states in which a purpose can be: not yet active, active, suspended, and terminated. 
Reasoning about the state of purposes introduces great expressiveness in \PurPL. 
For example, programmers can define a method which cannot be invoked unless a certain purpose is in a particular state. 
Furthermore, programmers can specify that the state of a purpose changed after the execution of a method. 
Let us consider the following code. 

{\small
\begin{lstlisting}[numbers=left,frame=single, 
xleftmargin=16pt,
caption={\strut}, label={lst:seven}]
void enroll [RI_Trial:active] (
    id: int {| RI_Trial, Press | rho1 |}, 
    obs_arg : int {| RI_Trial  | rho2 |})  {
    ... code ...
    }

void [RI_Trial:active] certRenewed 
     [RI_Trial:suspended] () {
    ... code ...
    RI_Trial.setState(active);
    }

... code, (*@\textit{Certification expires}@*)
RI_Trial.setState(suspended);

RI.enroll(P.id, Phys.obesity); (*@ \textcolor{red}{\XSolidBrush} @*)
\end{lstlisting}
}

Here, we are in the scenario in which the trial must be temporarily suspended until a required certification has been renewed. 
Line 14 changes the state of the purpose \texttt{RI\_Trial} to \texttt{suspended}. 
It does so with \PurPL's primitive operation \texttt{setState}.
In the meantime, we have refined the method declaration of \texttt{enroll} by adding the annotation \texttt{[RI\_Trial:active]} at line 1, after the name of the method and before the formla arguments. This notation specifies that such a purpose must be active for the method to be invoked. 
When we try to enroll our patient at line 16, the invocation is type rejected because the state of \texttt{RI\_Trial} is \texttt{suspended} rather than \texttt{active}. 

The code above also shows the declaration of the method \texttt{certRenewed} that is meant to be called after the certification has been renewed. \texttt{[RI\_Trial:suspended]} at line 8, which is after the name of the method and before the (empty sequence of) formal arguments,  states that \texttt{certRenewed} can be called only when \texttt{RI\_Trial} is suspended. Line 7 declares \texttt{[RI\_Trial:active]} before the name of the method and after the return type (\texttt{void}, here). It says that the state of \texttt{RI\_Trial} changes into \texttt{active} after the invocation. (Method \texttt{end\_trial} does not change the state of any purpose and so there is no use of this notation after \texttt{void} at line 1.)

Our program type checks successfully when we replace line 16 with the following two commands. 

{\small
\begin{lstlisting}[numbers=left,firstnumber=16,
xleftmargin=16pt,
frame=single]
certRenewed();
RI.enroll(P.id, P.obesity); (*@ \textcolor{dkgreen}{\CheckmarkBold} @*)
\end{lstlisting}
}

Notice that in order to type check successfully, the body of the methods \texttt{end\_trial} and \texttt{certRenewed} must justify the change of state for that purpose, for example using \PurPL's primitive operation \texttt{setState}, which changes the state of a purpose. 

\begin{figure}[b]
$
\begin{array}{rrl}
\textrm{Variables}  &x \in& \textsc{Var} \\
\textrm{Method Names} &m \in& \textsc{MethodNames} \\
\textrm{Field Names} &f \in& \textsc{FieldNames} \\
\textrm{Class Names} &C \in& \textsc{ClassNames} \\
%
%
\textrm{Ground Type} &\GroundType \ddefeq& \Bool \mid \Int \mid C  \\
\textrm{Purpose} &p \in& \textsc{Purpose} \\
\textrm{Purp. Variable} &\rho \in& \textsc{PurposeVar} \\
%
%
\textrm{Purpose set} &\pi \ddefeq&
  	\{\mid p_{1}, p_{2}, ...,p_{n}\mid\}
    \\
    &&
    \mid \{ \mid p_{1},...,p_{n} \mid \rho \mid\} \\
\textrm{Type}  &\Type \ddefeq&
  	\GroundType\,\pi \\
\textrm{Term} &t \ddefeq&
    \true \mid \false \mid n 
  	\mid x 
            \\
    &&
\mid t.f \mid t.m(\overline{x}) \mid \mathrm{new}_\pi\ C(\overline{t}) \\
\textrm{Value} &v \ddefeq&
  	\true \mid \false \mid n 
    \\&&\mid \mathrm{new}_\pi \ C(\overline{v}) \\
\textrm{Statements} &s \ddefeq&
  	x \defeq t \mid t.f \defeq t \mid s;s \mid \key{skip} \\
& & \mid \ite{t}{s}{s} \\ 
& & \mid \pwd{t}{s} \\ 
& & \mid x.\key{grant}(p)
\\ &&\mid x.\key{revoke}(p) \\
\textrm{Class Decl} &\CL \ddefeq&
  	\mathrm{class}\ C \ \mathrm{extends} \ D \{ \overline{\GroundType f},\overline{M} \} \\
\textrm{Method Decl} &M \ddefeq&
  	\Type \ m ( \overline{\forall \rho }. \overline{\Type x\Rightarrow \pi}) \{s\} \\
\textrm{Program} &prg \ddefeq&
  	\CL_{1},...,\CL_{n} 
\end{array}
$
  \caption{Syntax of \PurPL}\label{syntax}
\end{figure}

\section{Syntax and Semantics}\label{sec:syntax}

This section formalizes \PurPL~ in its version that omits the ability to specify states for purposes. Section \ref{sec:addingState} adds that ability to the formalism of this section.

\subsection{Syntax}

%
The syntax of \PurPL is in Fig. \ref{syntax}. This syntax closely relates to that of Featherweight Java and object calculi. 

The metavariable $C \in \textsc{ClassNames}$ ranges over class names, and we assume a top $\Obj \in \textsc{ClassNames}$. The metavariables 
$f$ and $m$ range over method and field names, respectively. We assume a finite set \textsc{Purpose} of purpose names ranged over by $p$, and a set of purpose variables \textsc{PurposeVar} ranged over by $\rho$. (We referred to these variables as \quoting{row variables} in Section \ref{sec:overview}.) A purpose set $\pi$ can be either a collection of purposes or a collection of purposes followed by a purpose variable $\rho$. 

A term (or expression) of \PurPL has type $\Type$ which consists of two elements. The first element is a ground type $\GroundType$: Integer, boolean, or the name of a class. 
The second element is the set of purposes for which it is available. (Ground types correspond to what Featherweight Java and most calculi simply call types. Our types carry purposes along, hence a distinct terminology.) 

A program \textit{prg} is a series of class declarations $\CL$. 
We assume that the computation starts with the execution of the method \texttt{main} of the class \texttt{Main}, which we assume is defined. A class declaration declares a class $C$ and which class $D$ it extends. Next, the declaration states the fields of the class and their ground type ($\overline{\GroundType \,f}$) and has a series of method declarations. 
A method declaration $M$ is of the form $\Type_{r} \ m ( \overline{\forall \rho }. \overline{\Type x\Rightarrow \pi}) \{s\}$. Here, $\Type_{r}$ is the return type of the method, and $s$ is the body of the method. The formal parameters of the method are prefixed with the quantification $\overline{\forall \rho}$ over the purpose variables which are encountered in their types. Each argument is declared with $\Type x\Rightarrow \pi$ where $x$ is the name of the argument, $\Type$ is its type (which contains a set of purposes), and $\pi$ is the new set of purposes that will be assigned to the type of the actual argument after the method is executed. 

Terms include integers, booleans, and common terms from Featherweight Java such as field access, method invocation, and the value constructor \key{new} for building a new object. 
To recall, the arguments of \key{new} are the values given to the fields of its class. The only difference, here, is that \key{new} carries the set of purposes $\pi$ for which the new object and its fields can be used. That is, fields of an object are not allowed to have different purposes in this formalism. 
(This choice may be easily relaxed but we did not see a reason to complicate our formalism for this less relevant aspect.)  
The values of the language are integers, booleans, and objects constructed with \key{new}. 
The syntax for method invocation is restricted. Since actual arguments may change their types after method invocation and our type system needs to keep track of that, we impose that only variables can be passed to a method. This imposes that arguments are first assigned to a variable before being passed to a method, a requirement that can be easily enforced with a compiler pass. 

Statements include assignments, field assignments, 
if-then-else, 
a while-loop, 
the sequence operation, and \texttt{skip}, as well as 
$x.\texttt{grant}(p)$, which adds $p$ to the set of purposes of $x$, and $x.\texttt{revoke}(p)$, which removes $p$ from that set. 
In our setting, these operations do not fail, i.e. \texttt{grant} does not fail if $p$ is already granted and \texttt{revoke} does not fail if $p$ is not in the set of purposes. 


\paragraph{Operational Semantics}
We omit showing the operational semantics of \PurPL. Once a \PurPL program type checks successfully, the purpose annotations in methods, \texttt{new}, as well as operations such as $\texttt{grant}(p)$ and $\texttt{revoke}(p)$ can be compiled away, as they were only useful for the type checker. Then, the operational semantics of a standard calculus with imperative objects can be used to provide a  dynamic semantics to \PurPL.


\begin{figure}
\[
\begin{array}{rrl}
\textrm{Method Type} &\MethodType \ddefeq&
  \overline{\forall \rho }. (\overline{\Type x\Rightarrow \pi} \to \Type) 
\end{array}
\]

  \begin{mathpar}
    \inferrule*[Lab=(Fields-Obj)]
    {
      ~
    }
    {
      \CT(\Obj) = \bullet
    }

    \inferrule*[Lab=(Fields-Class)]
    {
      \tc \classdecl{C}{D}\{\overline{\GroundType_C}\,\overline{f_C}, \overline{M}\} \\
      \CT(D) = \overline{\GroundType_D}\, \overline{f_D}
    }
    {
      \CT(C) = \overline{\GroundType_D}\, \overline{f_D},\ \overline{\GroundType_C}\, \overline{f_C}
    }


    \inferrule*[Lab=(Field-Type)]
    {      
      \GroundType\, f \in \CT(C) 
    }
    {
      \CT(C,f) = \GroundType
    }

    \inferrule*[Lab=(Method-Type-Present)]
    {
      \tc \classdecl{C}{D}\{\overline{\GroundType_C}\,\overline{f_C}, \overline{M}\} \\
      \Type_{r}\ m (\overline{\forall \rho }.\overline{\Type x\Rightarrow \pi}) \{s\} \in \overline{M}
    }
    {
      \CT(C,m) = \overline{\forall \rho }. (\overline{\Type x\Rightarrow \pi} \to \Type_{r})
    }

    \inferrule*[Lab=(Method-Type-Parent)]
    {
      \tc \classdecl{C}{D}\{\overline{\GroundType_C}\,\overline{f_C}, \overline{M}\} \\
      m \not\in \overline{M} \implies
      \CT(D,m) = \MethodType
    }
    {
      \CT(C,m) = \MethodType
    }
  \end{mathpar}
  \caption{Definition of $\CT(C)$, $\CT(C,f)$, and $\CT(C,m)$}\label{classTable}
\end{figure}

\subsection{Type System}

\paragraph{Auxiliary Functions for Class Declarations}


Our typing relations rely on auxiliary functions. 
Intuitively, they correspond to the various functions \textit{fields(C)}, \textit{mtype(m,C)}, and others of Featherweight Java, which are used to retrieve information from class declarations. 
As in Featherweight Java, our inference rules simply use these functions without passing them around explicitly, as well as retrieving class declarations with $\tc \textrm{class}~C~\textrm{extends}~D$.

%
We adopt the following notations: $CT(C)$, $CT(C,f)$, and $CT(C,m)$, which we describe below.  Their definition is in Fig. \ref{classTable}.  ($CT$ is meant to recall \quoting{Class Table}).


$CT(C)$ returns the sequence $\overline{\GroundType\, f}$ of field names with their ground types declared by the class $C$. Rules \textsc{(Fields-Obj)} and \textsc{(Fields-Class)} define this function, where $\bullet$ is the empty sequence. The latter rule retrieves the class definition of $C$ and returns its field declarations together with those inherited from  class $D$. 
As in Featherweight Java, we assume that duplicate names of fields and methods do not occur. 

$CT(C,f)$ retrieves the ground type of the field $f$ of class $C$. Rule \textsc{(Field-Type)} defines this function by simply retrieving all the fields and types of $C$ and looking therein for $f$. 

$CT(C,m)$ retrieves the signature of the method $m$ of class $C$. \textsc{(Method-Type-Present)} and \textsc{(Method-Type-Parent)} define this function. The former applies when $m$ is defined in $C$ while the latter retrieves the signature of $m$ from the extended class $D$ when $m$ is not defined in $C$. For the sake of clarity, $CT(C,m)$ does not return a method signature in the style that is closer to object-oriented notation $\Type_{r}\ m (\overline{\forall \rho }.\overline{\Type x\Rightarrow \pi}) \{s\}$ but is rearranged as a \emph{method type} of the form $\overline{\forall \rho }.(\overline{\Type x\Rightarrow \pi} \to \Type_{r})$. This is a more appropriate form also because purpose variables $\rho$s are bound in $\Type_{r}$.

\paragraph{Subtyping for Ground Types}
Our type system makes use of subtyping for ground types. This relation is the standard subtyping relation over types in object calculi and we do not show this relation here. 

The interaction between subtyping and purposes is rather trivial and it can be exemplified as follows. If a method requires an argument of type \textbf{Obj} with purposes \texttt{\{| p1,p2 | rho |\}}, we can pass a value of type, say, \textbf{Dog} (which extends \textbf{Obj}) that has been declared for purposes \texttt{\{| p1,p2,p3,p4 |\}}. First, we check \textbf{Dog} <: \textbf{Obj} with the ordinary subtyping relation and then we apply the match operation that we have discussed in Section \ref{sec:overview} to accommodate \texttt{\{| p1,p2,p3,p4 |\}} with the pattern \texttt{\{| p1,p2 | rho |\}}.



\begin{figure}
  \begin{mathpar}
  \begin{array}{l}
     \Gamma \tc \mathbf{true} : \Bool\,\pi \rtri \Gamma\\
     \Gamma \tc \mathbf{false} : \Bool\,\pi \rtri \Gamma\\
     \Gamma \tc n : \Int\,\pi \rtri \Gamma
     \end{array}

    \inferrule*
    {
      \Gamma(x) = \Type
    }
    {
      \Gamma \tc x : \Type \rtri \Gamma
    }\\

    \inferrule*[Lab=(Field-Access)]
    {
      \Gamma \tc t : C_t\pi_t \rtri \Gamma' \\
      \CT(C_t, f) = \GroundType
    }
    {
      \Gamma \tc t.f : \GroundType\pi_t \rtri \Gamma'
    }

       \inferrule*[Lab=(New)]
    {
      \Gamma_1 \tc t_1 : \GroundType_1\,\pi_1 \rtri \Gamma_2 \\\\
      \Gamma_2 \tc t_2 : \GroundType_2\,\pi_2 \rtri \Gamma_3 \\\\  
      \cdots\\\\
      \Gamma_{n} \tc t_n : \GroundType_n\,\pi_n \rtri \Gamma_{n+1} \\\\  
      \CT(C) = \overline{\GroundType'\, f} \\ \overline{\GroundType <: \GroundType'} \\\\
      \forall i, \pi \subseteq \pi_i \\
      \texttt{rhovars}(\pi) = \emptyset
    }
    {
      \Gamma_1 \tc \mathrm{new}_\pi C(\overline{t}) : C\pi \rtri \Gamma_{n+1}
    }

    \inferrule*[Lab=(Method-Invocation)]
    {
      \Gamma_0 \tc t : C_t\pi_t \rtri \Gamma_1 \\\\
      \Gamma_1 \tc x_1 : \GroundType_{1}\pi_{1} \rtri \Gamma_2 \\\\
            \Gamma_2 \tc x_2 : \GroundType_{2}\pi_{2} \rtri \Gamma_3 \\\\
            \cdots \\\\
      \Gamma_n \tc x_n : \GroundType_{n}\pi_{n} \rtri \Gamma_{n+1}\\\\
      \CT(C_t,m) = \forall \rho.\overline{(\GroundType' \pi' \Rightarrow \pi'')} \to \GroundType'_{\textrm{ret}}\pi \\
      \overline{\GroundType' <: \GroundType}\\\\
      \sigma = \bigcup_{x_i}\mathrm{pmatch}(\pi'_{i}, \pi_{i})\\
      \mathrm{is\mhyphen fn}(\sigma) \\
    }
    {
     \Gamma_0 \tc t.m(x_1\dots x_n) : \GroundType'_{\textrm{ret}}(\pi\sigma)
     \rtri \Gamma_{n+1}[x_i \mapsto \GroundType_{i}(\pi''_{i}\sigma)] }
  \end{mathpar}
  \caption{Term Typing}\label{termTyping}
\end{figure}

\paragraph{Term Typing}

Figure \ref{termTyping} shows the typing rules for terms. Our typing relation is of the form $\Gamma \tc t : \Type \rtri \Gamma'$, where $\Gamma$ and $\Gamma'$ are type environments, i.e., maps from variables to types $\Type$. 
The output type environment $\Gamma'$ states the new types for the variables of the term $t$ after having type checked it. Specifically, while variables do not change their ground types, they may change their set of purposes. 
%

The typing rules for integers, booleans, and variables in Fig. \ref{termTyping} are straightforward. To notice is that integers and booleans can be used for any set of purposes. %

Rule \textsc{(Field-Access)} type checks a field access $t.f$. 
We retrieve the type of $t$, whose ground type must be a class $C_t$, and has some set of purposes $\pi_t$. This means that all the fields of $C_t$ are declared for that set of purposes. 
Therefore, $\pi_t$ is used for the type of the overall field access. The ground type of the field is also retrieved from the class table with $CT(C_t,f)$. 

Rule \textsc{(New)} type checks the construction of a new object with $\mathrm{new}_\pi C(\overline{t})$ where $\pi$ is the set of purposes we intend for this new object. Rule \textsc{(New)} performs two checks. The first checks that the arguments of \key{new} are of the correct type. 
The types of the given arguments and those declared for the fields are compared with standard subtyping $<:$, as previously discussed. 
We retrieve the ground types declared for the field with $CT(C)$. Notice that arguments are evaluated (and therefore type checked) from left to right. In rule \textsc{(New)}, the output type environment is used as the input type environment for type checking the next argument. The second check determines that the arguments can be used for purposes $\pi$. To do so, we retrieve the purposes of each argument, i.e., $\pi_i$, and we check that they contain $\pi$. That is, those values have been declared for the purposes in $\pi$ and possibly for other purposes. In rule \textsc{(New)}, $\texttt{rhovars}(\pi)$ returns the set of purpose variables $\pi$. The rule checks that $\pi$ contains no purpose variables, i.e., it is made with concrete purposes only.  
Notice that notation $\pi \subseteq \pi_i$ trivially lifts the subset notion to purpose sets. 

Rule \textsc{(Method-Invocation)} type checks a method invocation $t.m(x_1\dots x_n)$. Here, $t$ must be type checked to be of some class $C_t$. 
We also type check all the arguments and retrieve their ground types and purpose sets. 
In the meantime, $CT(C,m)$ retrieves the signature of method $m$. Analogously to rule \textsc{(New)}, rule  \textsc{(Method-Invocation)} performs two checks. The first checks that the type of the actual arguments are subtypes of the type of formal arguments. The second checks purposes and this operation is complicated by the fact that both the purposes for the actual arguments and those of the formal arguments of $m$ can rely on purpose variables. 
We address this with a matching operation \key{pmatch} between two sets of purposes. 
Intuitively, $\key{pmatch}(\pi_1,\pi_2)$, for two sets of purposes $\pi_1$ and $\pi_2$, uses $\pi_1$ as a pattern to accommodate $\pi_2$ and, if $\pi_1$ makes use of a purpose variable $\rho$, then returns a substitution for $\rho$. Generally, a substitution $\sigma$ in \textsc{(Method-Invocation)} is a partial function from purpose variables to set of purposes. The \key{pmatch} operation, however, returns a singleton substitution that only assigns one purpose variable (the purpose variable that $\pi$ uses, if any) to a set of purposes. The \key{pmatch} operation is then used to match all the formal arguments of $m$ with the purposes of its actual arguments. The one-substitutions obtained are then combined to form a substitution $\sigma$ for all the purpose variables that are used in the signature of $m$. \texttt{is-fn} checks that $\sigma$ is a function. 
It is not a function when $\sigma$ contains two entries for the same purpose variable that maps it to two different purpose sets. It means that this purpose variable is used in two incompatible ways. 
The \key{pmatch} operation is defined as follows: (where $\emptyset$ is the empty substitution, and $\overline{p}$ denotes a finite sequence of purposes.)\\

\noindent\,\,\,$
{\mathrm{pmatch}(\{\mid p_1 \cdots p_k \mid \}, \{ \mid p_1 \cdots p_k \mid \}) = \emptyset}\\[2ex]
  {
  \begin{array}{c}
  \mathrm{pmatch}(\{\mid p_1 \cdots p_k ~|~ \rho \mid \}, \{ \mid p_1 \cdots p_k, p_m \cdots p_n \mid \}) \\[-0.5ex]
  = \\[-0.5ex]
  \{ \rho \mapsto \{ \mid  p_m \cdots p_n \mid  \}\}
  \end{array}
  }\\[2ex]
  {
  \begin{array}{c}
  \mathrm{pmatch}(\{ \mid \overline{p} \mid \rho_1 \mid \}, \{ \mid \overline{p}, p_m \cdots p_n \mid \rho_2  \mid \}) \\[-0.5ex]
  = \\[-0.5ex]
  \{ \rho_1 \mapsto \{ \mid  p_m \cdots p_n \mid \rho_2  \mid \}\}\\[0.5ex]
  \textit{Notice that }\mathrm{pmatch} \textit{ fails otherwise.}
  \end{array}
  }
$

\begin{figure}[h]
	\begin{mathpar}
    \inferrule*[Lab=(New)]
    {
      \forall i.\ \Gamma_i \tc t_i : G_i\,\pi_i \rtri \Gamma'_i \\
      \pi_i \subseteq \pi \\\\
      \Gamma_i = \Gamma'_{i-1} \\
      \Gamma_1 = \Gamma \\
      \CT(C) = \GroundType_1 \ldots \GroundType_n
    }
    {
      \Gamma \tc \mathrm{new}_\pi C(\overline{t}) : C\pi \rtri \Gamma'_n
    }

    \inferrule*[Lab=(Asgn)]
    {
      \Gamma \tc t: \GroundType\,\pi_t \rtri \Gamma' \\
      \Gamma'(x) = \GroundType\,\pi_x \\
      \pi_x \subseteq \pi_t
    }
    { 
     \Gamma\tc x \defeq t \rtri \Gamma' }

    \inferrule*[Lab=(Field-Update)]
    {
       \Gamma\tc t_1 : C\, \pi \rtri \Gamma' \\
      \CT(C, f) = \GroundType \\
       \Gamma'\tc t_2 : \GroundType\pi' \rtri \Gamma'' \\
      \pi \subseteq \pi'
    }
    { 
     \Gamma\tc t_1.f \defeq t_2 \rtri \Gamma'' }

    \inferrule*[Lab=(If-Then-Else)]
    {
       \Gamma\tc t : \Bool\ \pi \rtri \Gamma' \quad
       \Gamma' \tc s_1 \rtri \Gamma'_1 
      \quad
       \Gamma' \tc s_2 \rtri \Gamma'_2 \\
      \Gamma'' = \Gamma'_1 \sqcap \Gamma'_2 
    }
    { 
     \Gamma \tc \ite{t}{s_1}{s_2} \rtri \Gamma'' }

    \inferrule*[Lab=(Grant)]
    { \Gamma(x) = C\, \pi }
    { 
    \Gamma \tc x.\key{grant}(p) \rtri \Gamma[x \mapsto C(\pi \cup p)] }

    \inferrule*[Lab=(Revoke)]
    { 
      \Gamma(x) = C \,\pi 
    }
    { 
    \Gamma \tc x.\key{revoke}(p) \rtri \Gamma[x \mapsto C(\pi \setminus p)] }

     \inferrule*[Lab=(While)]
    {
			\Gamma \tc t : \Bool\ \pi \rtri \Gamma' \\ 
      \Gamma' \tc s \rtri \Gamma'' \\
      \Gamma''' = \Gamma' \sqcap \Gamma'' \\
      \Gamma''' \tc t : \Bool\ \pi' \rtri \Gamma''' \\
      \Gamma'''\tc s \rtri \Gamma'''
		}
		{ \Gamma \tc \pwd{t}{s}\rtri \Gamma''' }
	\end{mathpar}
	\caption{Statement Typing}\label{statementTyping}

\end{figure}

\paragraph{Statement Typing}

Figure \ref{statementTyping} shows the typing rules for statements. The typing relation has the form 
$\Gamma \tc s \rtri \Gamma'$. 

Rule \textsc{(Asgn)} type checks assignments $x := t$. 
We type check $t$ and we use the output type environment to retrieve the type of $x$. This is because the evaluation of $t$ might have modified the purposes of $x$. Then, we check that they are of the same type and that the purposes of the right-hand side contain those of $x$. 

\textsc{(Field-Update)} type checks field updates $t_1.f := t_2$. This rule works similarly to rule \textsc{(Asgn)} except that the type of the right-hand side is compared to that of the field $f$, which we retrieve with $CT(C,f)$. 

Rule \textsc{(Grant)} type checks a purpose grant operation $x.\key{grant}(p)$. This operation provides the new type environment to use in the rest of the program. This type environment has the purpose $p$ added to the purposes of $x$. 
Rule \textsc{(Revoke)} is similar and type checks a purpose revoke operation. The output type environment is such that the purpose $p$ is removed from the purposes of $x$.

Rule \textsc{(If-Then-Else)} type checks the if-then-else statement. We first type check the condition $t$, which must be of ground type $\Bool$. We use the output type environment $\Gamma'$ so obtained to type check both the then- and else-branch. The type checking of the then- and else-branch yields their corresponding output type environments $\Gamma_1'$ and $\Gamma_2'$, respectively. 
We compute the meet type environment of the two with $\Gamma_1' \sqcap \Gamma_2'$. This latter type environment is the one that we assign to the if-then-else. Intuitively, the meet of two type environments contains only the variables that occur in both type environments. Furthermore, for each entry $x:\GroundType_1 \, \pi_1$ of the first type environment and $x:\GroundType_2 \, \pi_2$ of the second, we have that the meet type environment contains $x:(\GroundType_1 \sqcap \GroundType_2) \, \pi_1 \sqcap \pi_2$, where the meet of ground type $\GroundType_1 \sqcap \GroundType_2$ here is the standard type meet operation over types in object calculi, and $\pi_1 \sqcap \pi_2$ is defined as follows. \\



\noindent$
\textit{Intersection } \overline{p} \cap \overline{p'} \textit{contains all $p$ in $\overline{p}$ and $\overline{p'}$.}\\[2ex]
\{ \mid \overline{p} \mid \} \sqcap \{ \mid \overline{p'} \mid \}  = \{ \mid \overline{p}\cap \overline{p'} \mid \} \\[2ex]
\{ \mid \overline{p} \mid \rho \mid \} \sqcap \{ \mid \overline{p'} \mid \} = \{ \mid \overline{p} \mid \} \sqcap \{ \mid \overline{p'} \mid \rho \mid \}  = \{ \mid \overline{p}\cap \overline{p'} \mid \} \\[0.5ex]
  \indent \textit{Notice that $\rho$ disappears.}\\[2ex]
\{ \mid \overline{p} \mid \rho \mid \} \sqcap \{ \mid \overline{p'} \mid \rho\mid \}  = \{ \mid \overline{p}\cap \overline{p'}\mid \rho \mid \} \\[0.5ex]
  \indent \textit{Notice that they need the same $\rho$.}\\
  \textit{$\sqcap$ fails otherwise.}\\
$


Rule \textsc{(While)} type checks the while statement. The difficulty here is that we need to type check the condition of the while and its body with a type environment that also covers for arbitrarily many consecutive runtime evaluations of such condition and body. We first type check the condition, we then type check the body. This covers only for the first evaluation of the condition and the first iteration of the body, but not, for example, a second evaluation of the condition. 
For example, the body of the \key{while} might revoke a purpose from a variable and the condition makes use of that variable exactly for that purpose. Then, a second evaluation of the condition at runtime is unsound. To address this, we take the output type environment after having type checked the condition and the one after having type checked the body and we compute the meet type environment. Then, we make sure that this meet type environment acts as a fixed point: This type environment must type check the condition without being modified, and must type check the body without being modified. This type environment is the output type environment after type checking the whole while statement. This is a conservative approach. For example, the evaluation of a condition may grant purposes $p_1$, $\ldots$, $p_{10}$ to variable $x$. 
In turn, the body of the \key{while}-statement may revoke purposes and leave $x$ with purposes $p_1$, $\ldots$, $p_{5}$. 
If the condition evaluates to $\false$, the code that follows this \key{while}-statement could, in principle, safely run with purposes $p_1$, $\ldots$, $p_{10}$ for $x$. However, our type system imposes that the code that follows is conservatively type checked knowing that $x$ has purposes $p_1$, $\ldots$, $p_{5}$ only, i.e., the meet purpose set. 

Typing rules for the sequence operator and \texttt{skip} are straightforward and not shown in Fig. \ref{statementTyping}. We also omit showing the typing rule for method declarations. 







\begin{figure}
\[
\begin{array}{rrl}
\textrm{Purpose State} &k \ddefeq&
  	\texttt{active} \\
    && \mid \texttt{notYetActive} \\
    && \mid \texttt{suspended} \\
    && \mid \texttt{terminated}  \\
\textrm{Statements} &s \ddefeq& \ldots 
\\
&& \mid p.\texttt{setState}(k)\\
\textrm{Purpose Env.} &\Pi \ddefeq&
  	\epsilon \mid p \mapsto k, \Pi \\
\textrm{Method Decl} &M \ddefeq&
  	(\Pi,\Type) \ m (\Pi, \overline{\forall \rho }. \overline{\Type x\Rightarrow \pi}) \{s\} \\
\textrm{Method Type} &\MethodType \ddefeq&
  \Pi, \overline{\forall \rho }. (\overline{\Type x\Rightarrow \pi} \to (\Pi,\Type)) 
\end{array}
\]
	\caption{Syntax of \PurPL with Purpose States}\label{syntaxState}
\end{figure}

\section{State of Purposes}\label{sec:addingState}

\begin{figure}
\begin{mathpar}
        \inferrule*[Lab=(setState)]
    {
    }
    { \Pi, \Gamma \tc p.\texttt{setState}(k) \rtri \Gamma, \Pi[p \mapsto k] }\\

      \inferrule*[Lab=(Method-Invocation)]
    {
      \Gamma_0 \tc t : C_t\pi_t \rtri \Gamma_1 \\\\
      \Gamma_1 \tc x_1 : \GroundType_{1}\pi_{1} \rtri \Gamma_2 \\\\
            \Gamma_2 \tc x_2 : \GroundType_{2}\pi_{2} \rtri \Gamma_3 \\\\
            \cdots \\\\
      \Gamma_n \tc x_n : \GroundType_{n}\pi_{n} \rtri \Gamma_{n+1}\\\\
      \CT(C_t,m) = \Pi_{m},\forall \rho.\overline{(\GroundType' \pi' \Rightarrow \pi'')} \to (\Pi'_{m},\GroundType'_{\textrm{ret}}\pi) \\
      \overline{\GroundType' <: \GroundType}\\\\
      \sigma = \bigcup_{x_i}\mathrm{pmatch}(\pi'_{i}, \pi_{i})\\
      \mathrm{is\mhyphen fn}(\sigma) \\\\
      \Pi_m \subseteq \Pi\\
      \Pi_{\textrm{new}} = \Pi + \Pi'_m
    }
    {
     \Pi,\Gamma_0 \tc t.m(\overline{x}) : \GroundType'_{\textrm{ret}}(\pi\sigma)
     \rtri \Gamma_{n+1}[x_i \mapsto \GroundType_{i}(\pi''_{i}\sigma)],\Pi_{\textrm{new}} }\\

    \inferrule*[Lab=(If-Then-Else)]
    {
       \Pi, \Gamma\tc t : \Bool\ \pi \rtri \Gamma',\Pi' \\\\
       \Pi', \Gamma' \tc s_1 \rtri \Gamma'_1, \Pi''  
      \\
       \Pi',\Gamma' \tc s_2 \rtri \Gamma'_2, \Pi''   \\\\
      \Gamma'' = \Gamma'_1 \sqcap \Gamma'_2 
    }
    { 
     \Pi, \Gamma \tc \ite{t}{s_1}{s_2} \rtri \Gamma'', \Pi''  }
  \end{mathpar}
  \caption{Relevant Typing Rules for Purpose State}\label{typingState}
\end{figure}

In this section, we augment the syntax and type system of the previous section with a treatment for the state of purposes. A purpose can be active, not yet activated, suspended, or terminated. The additions to the syntax of \PurPL are shown in Figure \ref{syntaxState}. 
In particular, we have added an operation $p.\texttt{setState}(k)$ that changes the state of a purpose into one of the mentioned states.

Method declarations can now specify that they can be invoked only so long that some purposes are in a particular state. Furthermore, they can specify that some purposes find themselves in some state after the execution of the method. Then, the syntax for method declarations $M$ and that of method types $\MethodType$ for the class table is updated as shown in Figure \ref{syntaxState}. They make use of a \emph{purpose environment} $\Pi$, which maps purposes to a purpose state. 

Our type system needs modifications in order to reason about this new information. Specifically, our term typing relation and statement typing relation also have a purpose environment as input. Furthermore, they also output a new purpose environment, as an operation might have changed the state of some purposes. The new typing relations for terms is 
$\Pi,
\Gamma \tc t :\Type \rtri \Gamma,\Pi$ and the new typing relation for statements is $\Pi,
\Gamma \tc s \rtri \Gamma,\Pi$. 

Most typing rules do not use the purpose environment and simply thread it through their rule. 
Figure \ref{typingState} shows the typing rules that are relevant to discuss. 

Rule \textsc{(setState)} type checks the \texttt{setState} statement. This operation simply modifies the current purpose state to record the new state for the purpose. 

Rule \textsc{(Method-invocation)} is modified to also check that the current states of the purposes $\Pi$ are appropriate for invoking the method. The requirements of method $m$ on the state of purposes are retrieved with $CT(C_t.m)$ and they are in $\Pi_{m}$ in the rule. Then, we simply check that $\Pi$ contains the entries of $\Pi_{m}$. The method type of $m$ also specifies in $\Pi'_{m}$ the new states that some purposes must have after the execution of the method. 
Then, the output purpose environment is $\Pi+\Pi'_{m}$, which is an update operation on purpose environments: It keeps the entries of $\Pi$ for those purposes that are not specified in $\Pi'_{m}$, it adds to $\Pi$ the entries of $\Pi'_{m}$ for purposes that are not specified in $\Pi$, and replaces the entries of $\Pi$ with those of $\Pi'_{m}$ for purposes that $\Pi$ and $\Pi'_{m}$ both specify.  

Rule \textsc{(If-Then-Else)} illustrates a complication. Consider the following code. 

{\small
\begin{lstlisting}[numbers=left,frame=single,
xleftmargin=15pt,caption={\strut}, label={lst:if}]
if flag
    then p.setState(suspended);
    else p.setState(active);

f();
\end{lstlisting}
}

We assume a boolean variable \texttt{flag} and a method \texttt{f} that requires the purpose \texttt{p} to be suspended when invoked.
What state should \texttt{p} be in after having type checked the if-then-else of lines 1-3? In other words, what purpose environment should we output with rule \textsc{(If-Then-Else)}? 
If we treat \texttt{p} as \texttt{suspended}, line 5 type checks successfully but a runtime execution may execute the else-branch, which makes that call unsound. 
If we treat \texttt{p} as \texttt{active}, line 5 is type rejected. This is an improvement but it is arbitrary as \texttt{f} could just as easily require \texttt{p} to be \texttt{active}. 
Our difficulty is that there is no \quoting{meet} state between \texttt{suspended} and \texttt{active}. 
Indeed, and more generally, states do not form a lattice. The solution that we adopt in \textsc{(If-Then-Else)} is that the then- and else-branch must return the same purpose environment. 
We acknowledge that this is restrictive and $\S$\ref{sec:future} discusses our future plan to handle this scenario with gradual typing. 
The \texttt{while}-statement, also, incurs in these same complications regarding the output purpose environment.



\section{Evaluation}\label{sec:evaluation}


We have implemented our \PurPL~ type system in Haskell. 
We have evaluated our type checker against the examples in Section \ref{sec:overview}. We confirm that our type checker repeats the analysis therein described, rejecting the examples in that section that should be rejected, and type checking successfully those that should be accepted. 


\paragraph{Experiments with Other Examples}
To test our type system, we developed other tests. 
For example, we have modeled a social media scenario where the user does not consent for their account to receive advertisements and type rejects attempts to show advertisements to it. 
In another example, we have modeled the scenario of a conference management that adopts a double blind policy for reviews and type rejects attempts to store authors' information into the fields of a paper object, while storing anonymous information makes the type checker succeed. 

\section{Limitations and Future Work}\label{sec:future}



\paragraph{Meta-theory of our Type System}

At the moment, our type system is not accompanied by  theorems and proofs that establish its meta-properties. In the future, we would like to prove safety theorems for our type system. Besides the progress and type preservation theorems, we would like to prove a \emph{compliance theorem} that says that, at runtime, data is not used for purposes other than those for which they have been declared.

\paragraph{Integrating Information Control-Flow} We presented our type system in its plain machinery that only reasons over purpose compliance. 
However, consider the following code and let us assume that \texttt{P.married} should not be used for the purpose \texttt{ToPublish}. 

{\small
\begin{lstlisting}[numbers=left,
xleftmargin=16pt,frame=single]
void f(flag : bool {| ToPublish |}) {
   ... code ...
}

myvar : bool {| ToPublish |} := true;
if (P.married)
    then { myvar := true; }
    else { myvar := false; }

f(myvar);
\end{lstlisting}
}

Although \texttt{P.married} is not passed to \texttt{f}, the if-then-else statement figures out its value and achieves the same effect as passing \texttt{P.married}. 
This issue suggests that our type system must be integrated with a control-flow analysis, as well. This is part of our future work.

\paragraph{Integrating Ownership} 

The type system must be augmented with a mechanism that enforces authorization constraints on operations such as \texttt{grant}/\texttt{revoke} and \texttt{setState}. For instance, in Listing~\ref{lst:three}, a hospital, acting upon a patient’s request, may legitimately grant the purpose \texttt{RI\_Trial} to \texttt{Phys.obesity}. In the absence of a sound authorization analysis, however, such grant/revoke operations could be invoked by arbitrary principals. Consequently, the research institute \texttt{RI} could illicitly execute \texttt{Phys.std.grant(RI\_Trial)}. Likewise, state transitions of a purpose must be performed only by its designated owner; for example, terminating 
\texttt{RI\_Trial} with \texttt{RI\_Trial.setState(terminated)} 
should be authorized exclusively by \texttt{RI}. A principled approach would be to track the authority of the execution context in which privileged operations occur and to enforce authorization constraints accordingly; under such a discipline, \texttt{RI\_Trial.setState(terminated)} could not be executed in the \texttt{Hospital} context. We leave the development of such an authorization mechanism to future work.


\paragraph{Strong Updates and Aliasing}

\,\!\PurPL adopts strong updates in which the type of a variable can change at runtime. Therefore, \PurPL~ suffers from well-known problems with aliasing.  
To make an example, consider the following code where the method \texttt{display} requires the purpose \texttt{ToPublish}. \\

{\small
\begin{lstlisting}[numbers=left,
xleftmargin=15pt,frame=single]
void f(a : bool {| ToPublish |}, 
        b : bool {| ToPublish |})  
    {
       a.revoke(ToPublish);
       display(b);
    }

myvar : bool {| ToPublish |} := true;
f(myvar,myvar);
\end{lstlisting}
}

This method is well-typed because \texttt{b} has the purpose \texttt{ToPublish} at the call \texttt{display(b)}. 
At runtime, however, \texttt{a.revoke(ToPublish)} has revoked such a purpose from \texttt{b}, and the call is unsound. 
This suggests that our type system must be integrated with aliases treatment, for example in the style of Rust or Garcia et al \cite{Garcia2014}. 

\paragraph{Gradual Typing for Purpose Sets} There are scenarios where the consent for using some data is simply known at runtime. 
For example, it may become known upon a click on a website or keyboard input from the user. 
Those situations cannot be currently modeled in \PurPL. We plan to handle them with gradual typing. Consider the following code, where \texttt{integer\_input()} reads an integer from the keyboard.

{\small
\begin{lstlisting}[numbers=left,
xleftmargin=15pt,frame=single]
void f(a : int {| ToPublish |}) 
    { ... code ... }    

myvar : int {| |} := 123; 
if integer_input() > 10 
    then myvar.grant(ToPublish);
    else skip;

f(myvar);

\end{lstlisting}
}

Our type system would type reject this program because, after having type checked the if-then-else, the purposes of \texttt{myvar} are the meet between \texttt{\{| |\}} (\quoting{no purpose}) and  \texttt{\{| ToPublish |\}}, which is \texttt{\{| |\}}. Line 10 then is type rejected. 
With gradual typing, however, we would replace line 5 with the following. 

{\small
\begin{lstlisting}[numbers=left,
xleftmargin=15pt,
firstnumber=5,frame=single]
myvar : int {| * |} := 123; 
\end{lstlisting}
}

This line makes use of (what we may call) the \emph{dynamic purpose set} $\star$, which states that the purpose set of \texttt{myvar} is unknown at compile-time and is discovered at runtime. At compile-time, the call \texttt{f(myvar)} type checks successfully, optimistically, but a runtime check is automatically inserted to verify that \texttt{myvar} possesses  \texttt{\{| ToPublish |\}} at runtime.
A common solution in gradual typing is that a compiler pass would insert a cast around \texttt{myvar} before passing it to \texttt{f}. That is, line 10 becomes 
{\small
\[
\texttt{f(myvar : int \{| * |\} => int \{| ToPublish\,|\})} 
\]
}

At runtime: If the then-branch at line 7 takes place, the cast would succeed and reduce to \texttt{myvar}, having the program execute \texttt{f(myvar)}. If the else-branch takes place then the cast would fail at runtime. 

Gradual typing can handle the scenario in Listing \ref{lst:if} (if-then-else and purpose state). 
When the then- and the else-branch do not agree for a state of a purpose, such a state can be set to (what we may call) the \emph{dynamic purpose state} $\star$, i.e., unknown at compile-time. The method call \texttt{f()} optimistically type checks but a runtime check is inserted before the call in order to check whether \texttt{p} has, at runtime, the purpose \texttt{suspended}.



\section{Related Work}\label{sec:related}

There are several works formalizing various privacy laws (e.g., HIPAA, GDPR, Right to erasure) using formal logics~\cite{alshugran, deyoung, robaldo} and enforcing them using static and dynamic techniques (e.g., runtime monitors)~\cite{chowdhury, rulekeeper, francois24,  tiwari}.  \citet{sok} has an elaborate taxonomy of rights and obligations enacted by modern Internet privacy and comprehensive
privacy laws as well as their enforcement. We encourage the reader to refer to the survey for a thorough treatment of the area. We focus on key enforcement techniques that are comparable to \PurPL. 

The Taint–Track–Control (TTC) approach and its web instantiation WebTTC provide a Privacy-by-Design architecture that enforces user-defined privacy policies by combining dynamic information-flow control with runtime verification~\cite{Hublet2024TTC}. In TTC, data is tainted with policy metadata and its propagation is monitored at runtime; candidate actions are reported to an external policy decision point, which may suppress or allow them to ensure compliance with temporal privacy policies. This design enables expressive user-specified policies and supports evolving consent at runtime, but enforcement occurs dynamically and requires instrumentation and monitoring overhead. WebTTC+ further extends this architecture with a privacy platform that mediates applications and enforcers to ensure compliance with privacy laws. 

Our work differs fundamentally in enforcement points and guarantees. Whereas TTC-style systems ensure compliance by runtime taint tracking and policy monitoring, we pursue a static, type-based approach in which the permitted purposes of data are encoded as typestate and checked at compile time. This shifts compliance reasoning earlier in the development lifecycle and enables developers to verify purpose usage without executing or instrumenting programs. Conceptually, TTC attaches dynamic policy tags to values, while our system treats the set of allowed purposes as part of the static type of data, enabling compile-time reasoning about purpose evolution and consent changes.


RuleKeeper~\cite{rulekeeper} is a GDPR-aware compliance framework where the data collected is attached to explicit purposes using a manifest file. Program dependence graphs are then analyzed for the data flow dependencies that violate the purpose policies specified in the manifest. Similar to RuleKeeper, \citet{Ferrara} investigates how taint analysis and backward slicing can be combined to enforce GDPR compliance.  RuleKeeper, however, allows  a gap between purposes declared in the manifest and the actual application; it relies on the programmer to ensure that the input manifest faithfully encodes the relation between variables and the corresponding purposes.  By contrast, \PurPL closes this gap by using purposes in types, allowing a more robust analysis.

Data Protection Language (DPL)~\cite{karami} is a language for GDPR enforcement. DPL attaches purpose-based contracts to objects and uses runtime monitoring to prevent  personal data from going to objects that implement no purpose or the wrong purpose. 
DPL is a runtime technique and comes with its own performance overhead. By contrast, \PurPL employs
a static approach. 

 \citet{bonatti} uses a logic-based description language to encode consent, business policies, and regulatory obligations; the compliance algorithms check that the business process comply with the policies.  By contrast, \PurPL works at the implementation level, ensuring that the purpose propagation is fully covered.

Our typestate approach is inspired  by Plaid and the work of Garcia et al. \cite{Wolff2011,Garcia2014}. 
Our system adapts the core mechanisms to a sophisticated setting that handles sets of purposes as states, treating them with row-polymorphism and, on top of that, our setting is such that our \quoting{states have a state themselves}. 


\paragraph{Semi-Formal Methods for Privacy Compliance.} 
\citet{elluri} develop a publicly-available knowledge graph of rules and regulations mandated for cloud data providers and consumers by GDPR.
rgdpOS~\cite{rgpdos} is a work in progress and aims at building a GDPR-aware operating system that enforces GDPR compliance at any software level. Crucially, it aims to have a kernel that ensures that functions only operate on personal data with matching purposes. It is noteworthy that the authors of rgdpOS acknowledge that the purpose matching is challenging and still unaddressed.

The DEFeND platform~\cite{defend} has been conceptualised around three axes of privacy
protection, i.e. Privacy By Design, Consent Management and Privacy Impact Assessment and Risk Management. 
It leverages encryption, authorisation, and anonymization to guarantee that the data is processed according to the data access rights. The architecture, however, lacks a formal approach. CookieBlock~\cite{bollinger} is a browser extension that uses machine learning to enforce GDPR cookie consent at the client. It automatically categorizes cookies by usage purpose using only the information provided in
the cookie itself. Software-Defined Data Protection~\cite{istvan}  decouples policies from storage node implementation (e.g., databases).  

The privacy guarantees offered by the above approaches are less formal and mostly use  defense-in-depth mechanisms for privacy enforcement. By contrast, \PurPL aims at offering sound privacy guarantees using principles grounded in formalism. Nevertheless, these works can help guide the development of rich features in \PurPL.

\paragraph{Information-flow Control (IFC)}  There is a whole body of work focusing on enforcing privacy policies using IFC~\cite{Denning1976Lattice, Volpano1996Sound, Sabelfeld2003Survey, Myers1997DLM, Austin2012JSFlow,  Stefan2012LIO, Yang2012Jeeves, Yang2015Relational, Arden2012FlowLimited}. IFC regulates how data propagates across program entities, ensuring that values labeled with higher confidentiality do not flow to less secure contexts. In contrast, purpose-oriented systems regulate why data may be used, enforcing that data is accessed only under authorized purposes.
Consequently, IFC enforces destination-based confidentiality (e.g., noninterference), whereas purpose systems enforce intent-based privacy constraints (e.g., purpose limitation). These dimensions are orthogonal: IFC cannot distinguish flows performed for different purposes, while purpose systems alone do not prevent unauthorized disclosure once a purpose is satisfied. Effective privacy enforcement therefore requires reasoning about both information flows and authorized purposes.

\section{Conclusion and Discussion}\label{sec:conclusion}

We have proposed a typestate approach for reasoning about the purpose of data. In this approach, sets of purposes act as the  state of a type. Our examples show that this use of typestate is a natural approach for reasoning about the usage of data w.r.t. their declared purposes. 
We presented a formal type system and we have reported on its implementation. 
Our type checker has been applied successfully to the examples in this paper and others, accepting programs or rejecting them when appropriate. 

In our experiments, we used the various \PurPL operations rather naturally, and purpose annotations have not been burdensome. However, a systematic study on the usability of \PurPL is part of our future work. 

This paper reports on our first step towards a practical programming language. 
We have identified in Section \ref{sec:future} several features that are 
essential if we intend \PurPL to be practical. 
The final type system that we will develop for \PurPL will 
need to simultaneously combine our typestate approach for purpose, information control-flow, ownership types, an aliasing treatment for soundness, and gradual typing. In this regard, one of the strengths of this paper is to present our approach for reasoning over purpose in isolation. 

\section{Acknowledgments}
Anitha Gollamudi was supported in part by the National Science Foundation under Grant No. NSF 2348304 and Office of Naval Research under Grant No. N000142512413. Samuel Dodson was supported in part by Draper Scholar Program. 
The work of Matteo Cimini and Joan Montas was supported in part by the National Science Foundation under Grant No. CCF-2317257. 





\bibliographystyle{ACM-Reference-Format}

\bibliography{references}

\end{document}